\documentclass{emulateapj}

\newcommand{\kms}{km s$^{-1}$}

\newcommand{\mbh}{M$_{\rm BH}$}

\newcommand{\hb}{H$\beta$}
\newcommand{\ha}{H$\alpha$}
\newcommand{\mt}{${\rm M}_{\rm T}$}
\newcommand{\mw}{${\rm M}_{\rm W}$}
\newcommand{\mn}{${\rm M}_{\rm N}$}
\newcommand{\mm}{${\rm M}_{\rm M}$}
\newcommand{\mk}{${\rm M}_{\rm K}$}
\newcommand{\mv}{${\rm M}_{\rm V}$}
\newcommand{\mvb}{${\rm M}_{{\rm V}\beta}$}
\newcommand{\mgc}{${\rm M}_{{\rm G}51}$}
\newcommand{\mgb}{${\rm M}_{{\rm G}\beta}$}
\newcommand{\mga}{${\rm M}_{{\rm G}\alpha}$}
\newcommand{\msa}{${\rm M}_{\rm Sa}$}
\newcommand{\msh}{${\rm M}_{\rm Sh}$}
\newcommand{\sline}{$\sigma_{\rm line}$}

\received{June 15 2007}
\accepted{October 9 2007}
\shorttitle{Black hole mass estimators for distant AGNs}
\shortauthors{McGill et al.}

\begin{document}

\title{Comparing and calibrating black hole mass estimators for
distant active galactic nuclei}

\author{Kathryn L. McGill\altaffilmark{1}, Jong-Hak Woo\altaffilmark{1,2},
Tommaso Treu\altaffilmark{1,3}, Matthew A. Malkan\altaffilmark{4}}

\altaffiltext{1}{Department of Physics, University of California,
Santa Barbara, CA 93106-9530; kmcgill@umail.ucsb.edu, woo@physics.ucsb.edu,
tt@physics.ucsb.edu} 
\altaffiltext{2}{Corresponding author}
\altaffiltext{3}{Alfred P. Sloan Research Fellow}
\altaffiltext{4}{Department of Physics and
Astronomy, University of California at Los Angeles, CA 90095-1547,
malkan@astro.ucla.edu} 

\begin{abstract}

Black hole mass (\mbh) is a fundamental property of active galactic
nuclei (AGNs). In the distant universe, \mbh\ is commonly estimated
using the MgII, \hb, or \ha\ emission line widths and the optical/UV
continuum or line luminosities, as proxies for the characteristic
velocity and size of the broad-line region. Although they all have a
common calibration in the local universe, a number of different
recipes are currently used in the literature. It is important to
verify the relative accuracy and consistency of the recipes, as
systematic changes could mimic evolutionary trends when comparing
various samples.  At $z=0.36$, all three lines can be observed at
optical wavelengths, providing a unique opportunity to compare
different empirical recipes.  We use spectra from the Keck Telescope
and the Sloan Digital Sky Survey to compare \mbh\ estimators for a
sample of nineteen AGNs at this redshift. We compare popular recipes
available from the literature, finding that \mbh\ estimates can differ
up to $0.38\pm0.05$ dex in the mean (or $0.13\pm0.05$ dex, if the same
virial coefficient is adopted). Finally, we provide a set of 30
internally self consistent recipes for determining \mbh\ from a
variety of observables. The intrinsic scatter between cross-calibrated
recipes is in the range $0.1-0.3$ dex. This should be considered as a
lower limit to the uncertainty of the \mbh\ estimators.

\end{abstract}
\keywords{black hole physics: accretion --- galaxies: active ---
galaxies: evolution --- quasars: general }

\section{Introduction}

Understanding the growth of supermassive black holes along with their
host galaxies is one of the fundamental questions in current
astrophysics \citep[e.g.][]{DMS05,Cro++06b}.  Black hole mass (\mbh) is a
key parameter in revealing the nature of black hole-galaxy coevolution
as well as the physics of active galactic nuclei (AGNs). However,
direct mass measurements using the motions of gas and stars in the
sphere of influence of a central black hole is limited to very nearby
galaxies \citep[e.g.][]{K+G01,F+F05}.

Beyond the very local universe, the so-called ``virial'' or
``empirically calibrated photo-ionization'' 
method based on the
reverberation sample is popularly used for active galaxies 
\citep[e.g.][]{WPM99, Kas++00, Kas++05, Ben++06a}.  This
method utilizes broad line widths as velocity indicators and
monochromatic continuum or line luminosities as indicators of
broad-line region size, hence estimating virial \mbh.
A combination of the MgII, \hb, or \ha\ broad emission line widths
and the 3000\AA, 5100\AA, \hb, or \ha\ luminosities is typically used,
depending on the redshift of the source and the observational setup.
Several equations have been presented in the literature to estimate
\mbh\ using various combinations of these indicators
\citep[e.g.,][]{W+U02b,W+U02a,M+J02,TMB04,Kol++06,G+H05,V+P06,Woo++06,Sal++07,N+T07,Tre++07}.

Although all three emission lines have a common calibration based on
the reverberation sample in the local universe, it is important to
verify that different recipes give consistent results; any systematic
changes could mimic evolutionary trends given that different recipes
are often used in various studies.

At $z=0.36$, all three lines can be observed at optical wavelengths,
providing a unique opportunity to cross-calibrate the different
methods of \mbh\ estimation. Using data from the Keck Telescope and the Sloan Digital Sky
Survey for a sample of nineteen AGNs at
$z=0.36$, we compare the different methods of estimating \mbh, and
derive a set of self-consistent equations for \mbh\ estimates using
every combination of velocity scale (FWHM and line dispersion \sline\,
of MgII, \hb, or \ha) and luminosity (3000\AA, 5100\AA - nuclear and
total - \hb, or \ha).

The paper is organized as follows.  In \S~\ref{sec:data} we describe
the sample selection, observations, and data reduction.  In
\S~\ref{sec:meas} we describe our line fitting process, based on expansion
in Gauss-Hermite series, and the resulting luminosity and width
measurements.  In \S~\ref{sec:form} we review the various formulae
adopted in the literature, and compare the various \mbh\ estimators.
In \S~\ref{sec:res} we present our self-consistent recipes.
Section~\ref{sec:sum} summarizes our results. 

Throughout this paper magnitudes are given in the AB scale. We assume
a concordance cosmology with matter and dark energy density
$\Omega_m=0.3$, $\Omega_{\Lambda}=0.7$, and Hubble constant H$_0$=70
kms$^{-1}$Mpc$^{-1}$.

\begin{figure*}
\epsscale{0.8}
\plotone{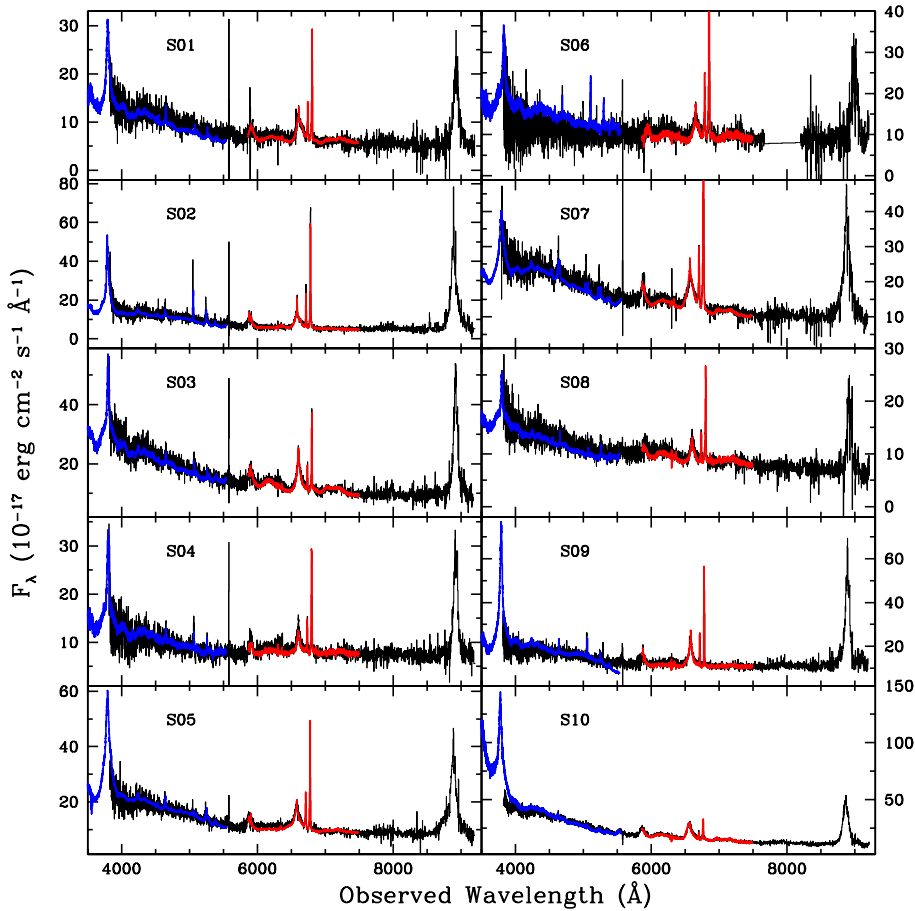}
\caption{Flux-calibrated spectra.  The SDSS spectra are shown in
black, and the Keck spectra are shown in blue and red.  The MgII line
can be seen on the far left of each wavelength range, while \hb\ is
located in the center and \ha\ to the far right.
}
\label{spectra1}
\end{figure*}

\begin{figure*}
\epsscale{0.8}
\plotone{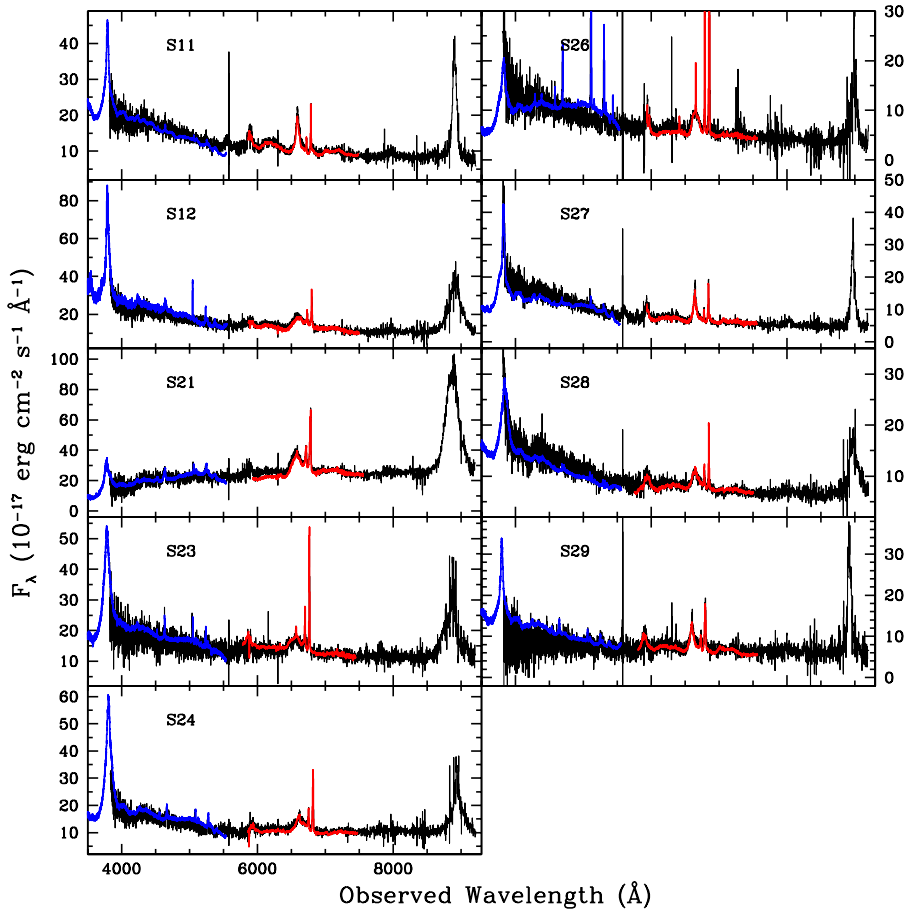}
\caption{As in Figure~\ref{spectra1} for objects S11 to S29.}
\label{spectra2}
\end{figure*}
\section{Sample selection, observations and data reduction}
\label{sec:data}

The AGN sample was initially selected for stellar velocity dispersion
($\sigma_{*}$) measurements to measure the M$_{\rm BH}$-$\sigma_{*}$
relation at $z=0.36$ \citep{TMB04,Woo++06}.  Readers are referred to the papers by
Woo et al.\ (2006) and Treu et al. (2007) -- where Keck red spectra
(5900\AA-7500\AA) and Hubble images of the sample were presented --
for more details. The relevant properties of the observed objects are
listed in Table~\ref{tab:sample}.

High signal-to-noise ratio spectra of nineteen targets were obtained with
the Low Resolution Imaging Spectrometer \citep[][hereafter
LRIS]{Oke++95} at the Keck-I telescope in five runs between March 2003
and July 2005, as detailed by \cite{Woo++06}. The red setup is
described by \cite{Woo++06}. In the blue, the 600 lines mm$^{-1}$
grism was used, yielding a pixel scale of 0.63\AA$\times 0\farcs$135
and a resolution of $\sim$145 km s$^{-1}$. Note that objects S16, S31,
and S99 included in the papers by \citet{Woo++06} or \citet{Tre++07}
lack Keck and/or SDSS spectra and are therefore not considered here.

The reduction of the blue spectra was very similar to that of the red
spectra described by \cite{Woo++06}, except that arc lamp emission
lines for Hg and Cd were used for wavelength calibration due to the
paucity of sky lines.  The flux was calibrated using
spectrophotometric stars or A0V type Hipparcos stars.  Galactic
extinction correction was applied to all data based on the average
extinction law derived in \citet{CCM89}.  The final step in the
reduction process involved normalizing all spectra to the proper AB
magnitude values from the Sloan photometric database. This was
achieved by calculating synthetic $g'$ and $r'$ magnitudes from the
spectra taking into account the SDSS bandpasses and finding the
constant multiplicative factor appropriate to match the SDSS
photometry.  The final flux-calibrated spectra of all nineteen observed AGNs
are shown in Figures~\ref{spectra1} and~\ref{spectra2}.

\begin{figure}
\epsscale{1.0}
\plotone{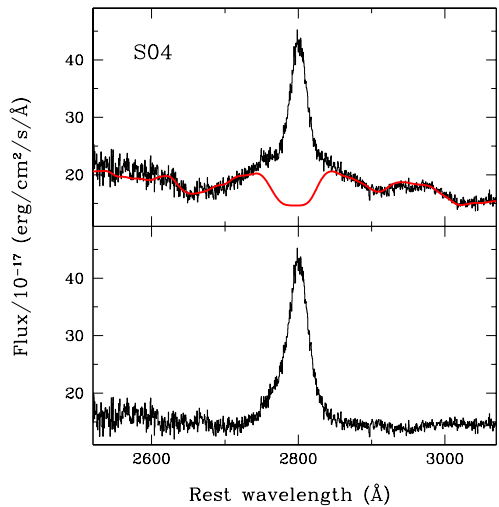}
\caption{Example of nuclear Fe subtraction. The top panel shows a typical observed spectrum (black histogram), together with the nuclear iron emission template matched in intensity and resolution (red line). The bottom line shows the residuals after Fe subtraction. Note that the bump redwards of MgII emission has disappeared in the residuals.}
\label{fig:FeII}
\end{figure}

\section{Measurements}
\label{sec:meas}

The relevant quantities for our purpose are line widths and line and
continuum fluxes. In this paper, line widths are measured as FWHM or
line dispersion, i.e. the square root of the second central moment:

\begin{equation}
\sigma_{\rm line}^2=\frac{\int ({\rm F}_{\lambda}-{\rm C}) (\lambda-\lambda_0)^2 d\lambda}{\int ({\rm F}_{\lambda}-\rm C ) d\lambda}
\end{equation}

where F$_{\lambda}$ is the flux density, C is the continuum, $\lambda$
is the wavelength and $\lambda_0$ is the central wavelength of the
line. As described in this section, we derive line widths and fluxes
by fitting a Gauss-Hermite series to the data after subtraction of the
continuum, Fe emission, and when necessary a narrow component of the line.
While the Gauss-Hermite fit is not necessary for
the high S/N ratio Keck data -- where the quantities can be measured
directly from the data \citep{TMB04,Woo++06} -- the fitting
procedure is needed to obtain robust measurements for the noisier SDSS
data. As we demonstrate below by comparing the H$\beta$ measurements
from Keck and SDSS data, this fitting procedure gives consistent
results between the two datasets, and therefore indicates that the
quality of the SDSS data is adequate to measure the width and flux of
H$\beta$ and, especially, H$\alpha$, since the latter is considerably
stronger.

\subsection{Gauss-Hermite Fitting}
\label{ssec:gauss}

The emission lines, especially for \hb, are often asymmetrical, thus
making a symmetrical Gaussian approximation of the line profiles
undesirable.  To account for the asymmetries in the emission lines, we
fit a truncated Gauss-Hermite series to the profiles \citep{v+F93}.
The main advantage of the Gauss-Hermite expansion is that it provides
an orthonormal basis set and that the coefficients of the Hermite
polynomials (commonly referred to as $h_3$, $h_4$, etc.) can be
derived by straightforward linear minimization, leaving only two
non-linear parameters (the center and the width of the Gaussian). Furthermore,
the coefficients can be interpreted in terms of the kinematics of the
tracing population \citep[e.g.][]{Ger93}. The best fit profiles are
then used to measure the luminosity, FWHM, and \sline\, of the
emission lines as parameters for the \mbh\ formulae.

We begin the fitting process with continuum subtraction.  We identify
the continuum level on each side of the three emission lines, using a
window of 60 $\rm \AA$. In the case of MgII, a narrower 40 $\rm \AA$
window is used because the blue continuum of MgII is close to the end
of the observed spectral range.  For MgII, the ranges typically used
are 2660-2700 $\rm \AA$ for the blue continuum level and 2930-2970
$\rm \AA$ for the red continuum level.  Similarly for \hb, the ranges
are 4670-4730 \AA\, and 5080-5140 \AA\ for blue and red, respectively,
and for \ha\ the ranges are 6290-6350 $\rm \AA$ and 6700-6760 $\rm
\AA$.  We then independently subtract the continuum by linear
interpolation for each line.

\begin{figure}
\epsscale{1.2}
\plotone{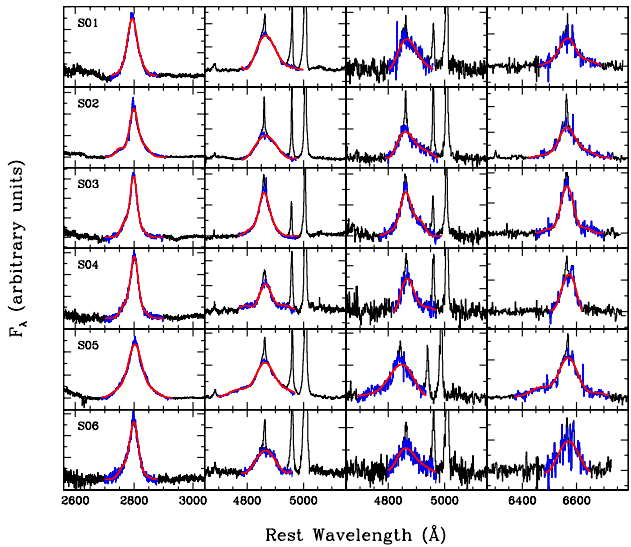}
\caption{Gauss-Hermite broad line fits.  From left to right, the boxes
show the fits for the Keck MgII line, Keck H$\beta$ line, SDSS
H$\beta$ line, and SDSS H$\alpha$ line for each object. The
continuum-subtracted line is shown in black, the broad component is
shown in blue, and the broad-line fit is shown in red.
}
\label{fig:fits1}
\end{figure}

\begin{figure}
\epsscale{1.2}
\plotone{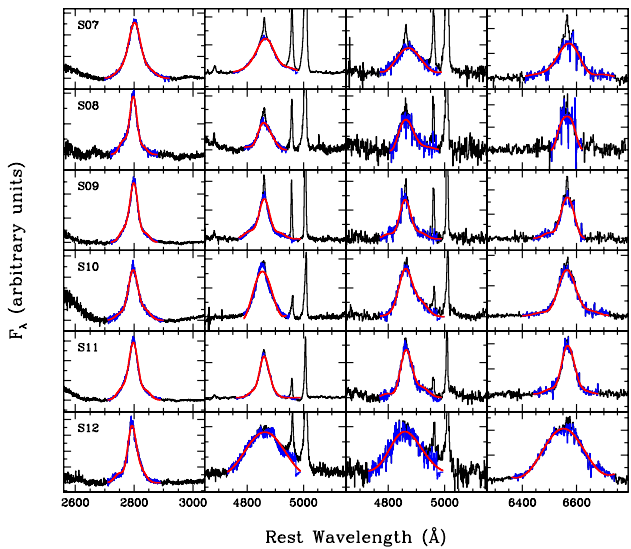}
\caption{As in Figure~\ref{fig:fits1} for objects S07 to S12.}
\label{fig:fits2}
\end{figure}

\begin{figure}
\epsscale{1.2}
\plotone{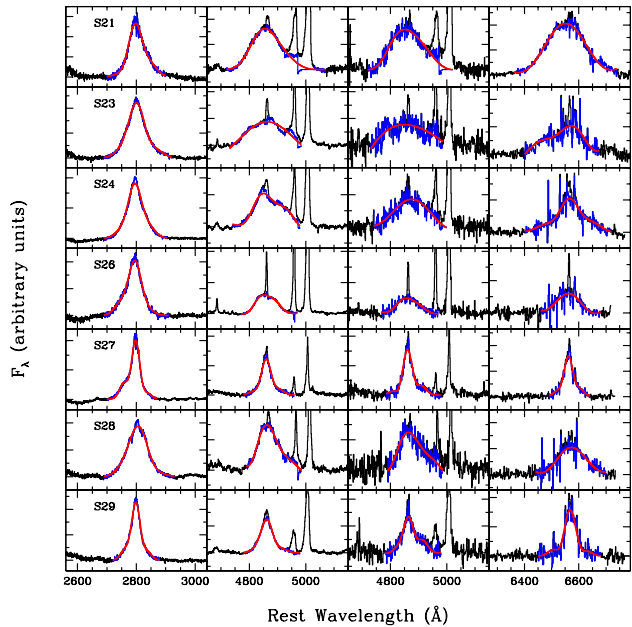}
\caption{As in Figure~\ref{fig:fits1} for objects S21 to S29.}
\label{fig:fits3}
\end{figure}
Together with the featureless continuum, we also remove broad nuclear
Fe emission underneath MgII and H$\beta$. For this task we use
template spectra of I Zw 1 kindly provided by Todd Boroson and Ross
McLure.  The procedure is similar to that described by
\citet{Woo++06}. A typical example of Fe subtraction in the wavelength
region around MgII is shown in Figure~\ref{fig:FeII}. Removing nuclear
Fe emission changes the measured width of H$\beta$ by a negligible
amount (the FWHM is unchanged and \sline\, is reduced by 0.013
dex), but it has a significant effect on MgII (the FWHM is reduced by
0.027 dex, while \sline\, is reduced by 0.106 dex).

Before fitting the broad lines, we subtract out the narrow lines,
extending the procedure described by \citet{Woo++06}.  For \hb\, this
involves subtraction of the [OIII] $\lambda$4959 and $\lambda$5007
narrow lines, as well as the narrow \hb\ line.  We subtract [OIII]
$\lambda$5007 directly, and subtract [OIII] $\lambda$4959 by dividing
[OIII] $\lambda$5007 by 3 and blueshifting.  The narrow component of
\hb\ is subtracted by rescaling and blueshifting [OIII] $\lambda$5007.
The line ratio H$\beta_{\rm narrow}$ / [\ion{O}{3}]$\lambda$5007 was
allowed to range between 1/20 and the maximum value consistent with
the absence of ``dips'' in the broad component (typically 1/10-1/7;
e.g. Marziani et al.\ 2003). The adopted value of the scale factor is
listed in Table~\ref{tab:derive}. The narrow component of \ha\ is
subtracted by multiplying the determined \hb\ narrow component by 3.1
\citep{Mal83,Ost89} and redshifting.

No attempt is made to remove narrow NII emission lines around
H$\alpha$, since they are effectively rejected by the Gauss-Hermite
fitting procedure as noise spikes, nor the narrow component (if
present) of MgII.

The final step in approximating the line profiles involves fitting the
Gauss-Hermite series to the broad lines.  The fitting procedure finds
the minimum $\chi^2$ by increasing the order of the Hermite
polynomials as required by the data (i.e. only if the goodness of fit,
measured by the reduced $\chi^2$, improves).  Most objects required up
to order 6 polynomials (i.e. $h_6$) to plateau in reduced $\chi^2$.
Figures~\ref{fig:fits1} to~\ref{fig:fits3} show our best fits to the MgII,
H$\beta$, and H$\alpha$ lines for the nineteen objects. The derived
measurements of \sline\, and FWHM, after removal of the instrumental
resolution, are listed in Table~\ref{tab:derive}.

\subsection{Luminosities}
\label{ssec:lum}

The formulae for \mbh\ estimates require line luminosities or
monochromatic continuum luminosities at given wavelengths. For
consistency with \citet{G+H05} the line luminosities are calculated
from the total flux for the combined broad and narrow components of
the \hb\, and \ha\, emission lines, where the broad component is taken
as the best Gauss-Hermite fit, and the narrow component is taken as
the appropriately scaled and shifted [OIII] $\lambda$5007 narrow line
(see \S~\ref{ssec:gauss}). We note that the narrow components
contribute only a small fraction of the total flux of the Balmer lines
(e.g., $\sim$5\% for H$\beta$ -- see Table~2, and similarly to H$\alpha$) and
therefore they make a contribution of about 0.01 dex to the broad-line
size estimates.  The H$\alpha$ to H$\beta$ flux ratio ranges between 3
and 7 as expected for Seyfert 1s
\cite[e.g.][]{Lac++82}. 

The total continuum luminosity at 3000 (5100) $\rm \AA$ is calculated
from the average flux in the 2950-3050 (5050-5150) $\rm \AA$ rest
frame. In this paper, we use the term total continuum luminosity to
indicate the total luminosity as measured within the spectroscopic
aperture, i.e. without removing the host galaxy
contamination. The fraction of host galaxy contamination
depends on the properties of the individual object as well as on the
instrumental setup, and it is hence a source of scatter. Nevertheless, 
the total luminosity is often the only measurement available and 
therefore it is important to investigate estimators based on this quantity.

The total continuum luminosities at 5100 $\rm \AA$ measured from the
SDSS spectra agree to within a few per cent of those inferred from the
Keck spectra and those listed in the paper by \citet{Woo++06}. By
comparing the values listed here with respect to those given by
\citet{Woo++06}, we infer 0.014 dex as the error on the continuum
(L$_{3000}$ and L$_{5100}$).  For line luminosities, we compare
measurements on the fit with measurements on the data, and we take the
r.m.s. scatter as the average error. This error is 0.011 and 0.062 dex
on L$_{\rm H\beta}$ and L$_{\rm H\alpha}$, respectively.  Nuclear
luminosities are taken from the HST measurements presented by
\citet{Tre++07}, except for three objects (S11, S28, and S29), where
the nuclear luminosity has been estimated from scaling the total
luminosity by the average nuclear fraction for the sample, 0.31.  In
addition to measurement errors, AGN variability effectively limits the
accuracy of the calibration of luminosity-based estimators, to the
typical level of variability of 5-10\% \citep[e.g.,][]{W+M00,Woo++07}.

\begin{figure}
\epsscale{1.0}
\plotone{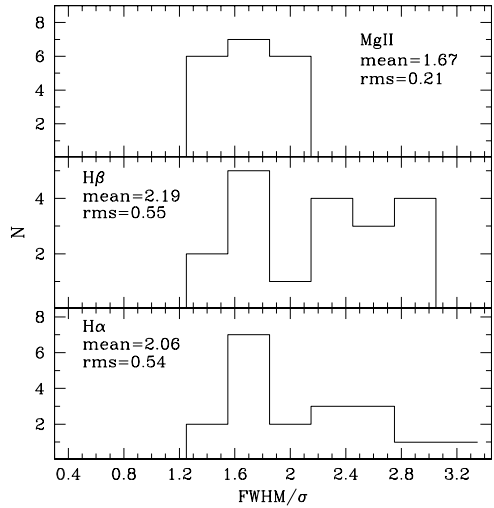}
\caption{Distribution of FWHM to \sline\, ratios for MgII, H$\beta$, and
H$\alpha$. For comparison, the expected value for a Gaussian is
2.35. The typical errors on the measurements are 0.03 dex for MgII and
\hb\, and 0.06 dex for \ha.}
\label{fig:widthhisto}
\end{figure}

\subsection{Line Widths}
\label{ssec:width}

The \mbh\ estimators depend either on \sline\, or on the FWHM as a
measurement of line width, and hence of the broad line kinematics.
Both of these quantities were measured from the Gauss-Hermite fit to
the broad lines. Errors on the Keck line width measurements are
obtained by comparison with those reported by \citet{Woo++06}, which
were measured independently and did not rely on Gauss-Hermite
expansion. The average error is 0.017 dex for the MgII and \hb\ FWHM and
\sline. The errors on the SDSS \ha\, line width measurements are
obtained by comparing the results for \hb\, with the Keck fit and
assuming that the error is the same for \ha\, and \hb. The average
error is 0.051 dex for \sline\, and 0.040 dex for the FWHM.

\begin{figure}
\epsscale{1.0}
\plotone{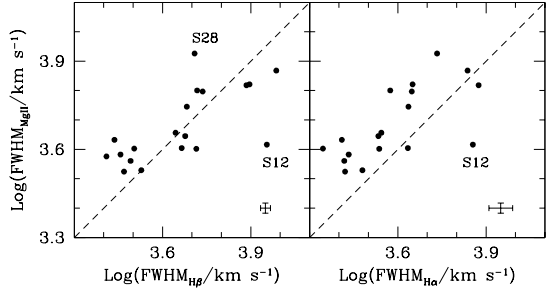}
\caption{Comparison of the width of MgII with that of H$\beta$ and
that of H$\alpha$.
The average errors are presented in the bottom right conner. The two
objects furthest away from the ratio=1 line (dashed line) are labeled
for easy comparison with the line profiles shown in
Figures~\ref{fig:fits2} and~\ref{fig:fits3}.}
\label{fig:widthcomp}
\section{Review of optical-UV \mbh\ Estimators}
\label{sec:form}
\end{figure}

The distribution of the FWHM to \sline\, ratios is shown in
Figure~\ref{fig:widthhisto}. The average ratio for H$\beta$ and
H$\alpha$ is close to the Gaussian value (2.35), although with large
scatter, consistent with the sample of \citet{Pet++04}. For MgII the
average is considerably smaller indicating significant departure from
Gaussianity. A large range of FWHM/\sline\, ratios indicates a large
scatter between \mbh\ based on FWHM and \mbh\ based on \sline\ for the
sample since velocity is simply derived either from FWHM or \sline\
by multiplying by a constant as shown in \S~\ref{sec:form} (see
detailed discussion by \citet{Col++06}).

Figure~\ref{fig:widthcomp} compares the FWHM of MgII with that of
H$\beta$ and H$\alpha$. The FWHM are correlated albeit with
substantial scatter. Comparing all the velocity scales, the average
ratios and r.m.s. scatters (in parenthesis) are: 
$\langle \log({\rm FWHM_{MgII}/FWHM_{H\beta}}) \rangle =0.02\pm0.03 (0.13)$, 
$\langle \log({\rm FWHM_{H\beta}/FWHM_{H\alpha}}) \rangle =0.09\pm0.02 (0.07)$, 
$\langle \log(\sigma_{\rm MgII}/\sigma_{\rm H\beta}) \rangle =0.13\pm0.02 (0.10)$, 
$\langle \log(\sigma_{\rm H\beta}/\sigma_{\rm H\alpha}) \rangle =0.07\pm0.02 (0.10)$. 

Summarizing these relations, the width of H$\alpha$ is generally
narrower by $\sim$20\% than that of H$\beta$ as expected from other
studies \citep[e.g.][]{Shu84,G+H05}, while MgII and \hb\, are similar
in FWHM but not in \sline. These differences are expected given that
different lines trace different species and that their shapes reflect
the ionization and excitation variations throughout the region. This
finding implies that each line has to be calibrated independently as a 
velocity estimator and that one cannot go from FWHM to \sline\,
using the simple scaling for a Gaussian distribution. Therefore, we
conclude that in general \ion{Mg}{2} and Balmer line width cannot be
used interchangeably, although in the case of the FWHM of Mg and
\hb\, the ratio is close to unity \citep[see][for further
discussion]{M+J02,Sal++07}.

As far as \mbh\, estimators are concerned, the scatter is of order 0.1 dex
(i.e. significantly larger than the measurement errors), which sets a
lower limit of $\sim 0.1-0.2$ dex on the relative uncertainty of the
cross calibration of simple velocity estimators based on line widths.

We begin this section with a list of the 12 formulae for \mbh\
estimation considered in this paper (\S~\ref{ssec:formi}). In these
formulae, we adopt a notation where L$_{5100,t}=$total luminosity
$\lambda L_{\lambda}$ at $\lambda=5100\rm \AA$, L$_{5100,n}=$nuclear
luminosity $\lambda L_{\lambda}$ at $\lambda=5100\rm \AA$, and
L$_{3000}=\lambda L_{\lambda}$ at $\lambda=3000\rm \AA$. All formulae
are given in the original notation, without applying any correction
for different assumptions on the virial coefficient.

In \S~\ref{ssec:res} we will compare the various estimators to infer
how much they differ when applied to the same sample of objects before
presenting our cross-calibrated recipes in \S~\ref{sec:res}.

\subsection{Summary of existing recipes}
\label{ssec:formi}

\citet{M+J02}, \citet{Kol++06} and \citet{Sal++07} give equations
based on the width of the MgII line and the optical/UV continuum
luminosity:

\begin{equation}
M_{\rm M} = 3.37 \left(L_{\rm 3000} \over 10^{37} \mbox{ \rm W } \right)^{0.47} \left(\rm FWHM_{\rm MgII} \over \mbox{ \rm km s}^{-1} \right)^{2} M_{\odot},
\label{eq:MM}
\end{equation}

\begin{equation}
M_{\rm K} = 2.04 \left(L_{\rm 3000} \over 10^{44} \mbox{ \rm erg} s^{-1} \right)^{0.88} \left(\rm FWHM_{\rm MgII} \over \mbox{ \rm km s}^{-1} \right)^{2} M_{\odot},
\label{eq:MK}
\end{equation}

\begin{equation}
M_{\rm Sa} = 10^{7.69} \left(L_{5100,t} \over 10^{44} \mbox{ \rm erg s}^{-1} \right)^{0.5} \left(\rm FWHM_{\rm MgII} \over 3000 \mbox{ \rm km s}^{-1} \right)^{2} M_{\odot}.
\label{eq:MSa}
\end{equation}

\noindent
\citet{G+H05} and \citet{V+P06} present equations based on the width
and luminosities of the \hb\ and \ha\, broad lines:

\begin{equation}
M_{\rm G\beta} = 3.6 \times 10^{6} \left(L_{\rm H\beta} \over 10^{42} \mbox{ \rm erg s}^{-1} \right)^{0.56} \left(\rm FWHM_{\rm H\beta} \over 1000 \mbox{ \rm km s}^{-1} \right)^{2} M_{\odot},
\label{eq:MGB}
\end{equation}

\begin{equation}
M_{\rm G\alpha} = 2.0 \times 10^{6} \left({L_{\rm H\alpha} \over 10^{42} \mbox{ \rm erg s}^{-1}} \right)^{0.55} \left(\rm FWHM_{\rm H\alpha} \over 1000 \mbox{ \rm km s}^{-1} \right)^{2.06} M_{\odot},
\label{eq:MGA}
\end{equation}

\begin{equation}
M_{\rm V\beta} = 10^{6.67} \left(L_{{\rm H}\beta} \over 10^{42} \mbox{ \rm erg s}^{-1} \right)^{0.63} \left(\rm FWHM_{\rm H\beta} \over 1000 \mbox{ \rm km s}^{-1} \right)^{2} M_{\odot}.
\label{eq:MVb}
\end{equation}

\noindent
\citet{Shi++03}, \citet{G+H05}, \citet{V+P06}, \citet{Woo++06},
\citet{N+T07}, and \citet{Tre++07} adopt the following formulae based
on L$_{5100}$ and the width of the \hb\ broad line:

\begin{equation}
M_{\rm Sh} = 10^{7.69} \left(L_{5100,t} \over 10^{44} \mbox{ \rm erg s}^{-1} \right)^{0.5} \left(\rm FWHM_{\rm H\beta} \over 3000 \mbox{ \rm km s}^{-1} \right)^{2} M_{\odot},
\label{eq:MSh}
\end{equation}

\begin{equation}
M_{\rm G51}=4.4 \times 10^{6} \left(L_{5100,n} \over 10^{44} \mbox{ \rm erg s}^{-1} \right)^{0.64} \left(\rm FWHM_{\rm H\beta} \over 1000 \mbox{ \rm km s}^{-1} \right)^{2} M_{\odot},
\label{eq:MG51}
\end{equation}

\begin{equation}
M_{\rm V} = 10^{6.91} \left(L_{5100,t} \over 10^{44} \mbox{ \rm erg s}^{-1} \right)^{0.5} \left(\rm FWHM_{\rm H\beta} \over 1000 \mbox{ \rm km s}^{-1} \right)^{2} M_{\odot},
\label{eq:MV}
\end{equation}

\begin{equation}
M_{\rm W} = 2.15 \times 10^{8} \left(L_{\rm 5100,t} \over 10^{44} \mbox{ \rm erg s}^{-1} \right)^
{0.69}  \left(\sigma_{\rm H\beta} \over 3000 \mbox{ \rm km s}^{-1} \right)^{2} M_{\odot},
\label{eq:MW}
\end{equation}

\begin{equation}
M_{\rm N} = 1.05 \times 10^{8} \left(L_{5100,t} \over 10^{46} \mbox{ \rm erg s}^{-1} \right)^{0.65} \left(\rm FWHM_{H\beta} \over 1000 \mbox{ \rm km s}^{-1} \right)^{2} M_{\odot},
\label{eq:MN}
\end{equation}

\begin{equation}
M_{\rm T} = 10^{8.58} \left(L_{5100,n} \over 10^{44} \mbox{ \rm erg s}^{-1} \right)^{0.518} \left(\sigma_{\rm H\beta} \over 3000 \mbox{ \rm km s}^{-1} \right)^{2} M_{\odot}.
\label{eq:MT}
\end{equation}

\noindent
Note that the formulae listed above adopt different estimators of broad-line 
region velocity (\sline\, or FWHM) and size (continuum
luminosity or line luminosity).

\begin{figure*}
\epsscale{0.8}
\plotone{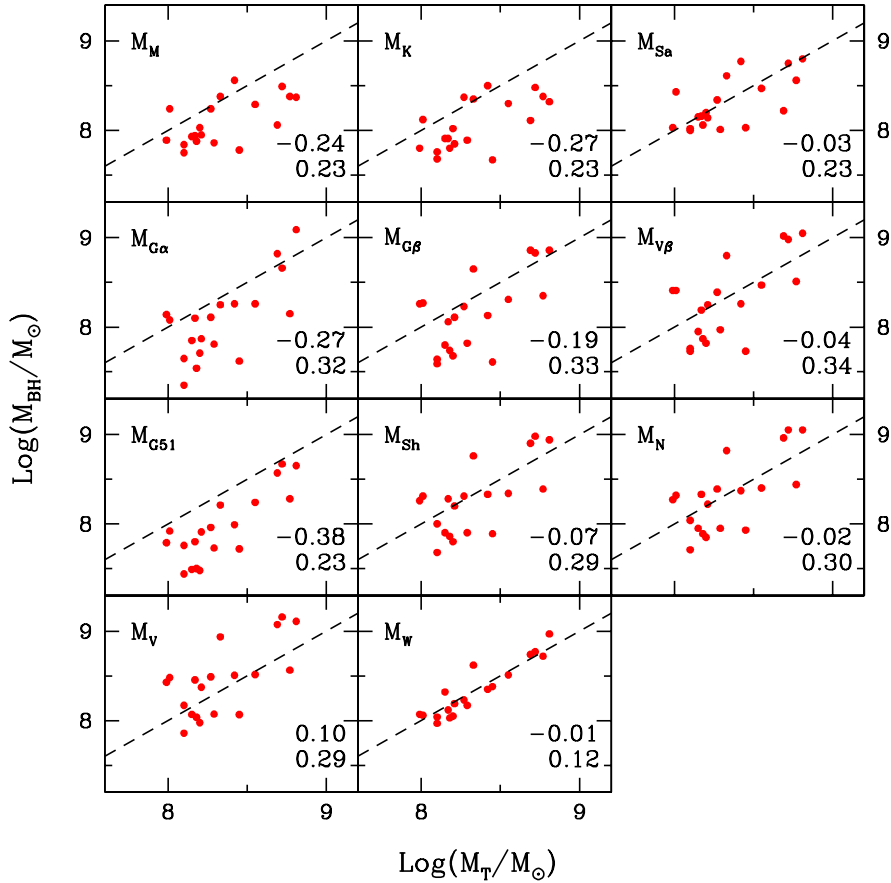}
\caption{Comparison of the consistency of various estimators adopted
in the literature, as summarized in Section~\ref{sec:form}. The
average difference ($\log M_{\rm BH} - \log$ \mt) and r.m.s. scatter in
dex are listed in the bottom right corner. The formulae are taken
directly from the literature with no adjustments for the
difference in the choice of virial coefficients. 
}
\label{fig:formulas}
\end{figure*}

\subsection{Comparison of \mbh\, estimators}
\label{ssec:res}

In this section we assess the relative consistency of the \mbh\
estimators taken from the literature by comparison with our fiducial
black hole mass \mt\, (as listed in the paper by Treu et al.\ 2007).
The choice of the fiducial estimator is based on reverberation mapping
studies of local AGNs, which are mostly based on H$\beta$ and
$L_{5100}$. These studies show that \sline\, is the most robust
velocity estimator \citep{Pet++04,Col++06}. They determine the slope
of the size-luminosity relation, taking into account the host galaxy
contamination \citep{Ben++06a}, and they set the virial coefficient by
requiring that the \mbh-$\sigma_*$ relation be the same for active and
quiescent galaxies, modulo selection effects \citep{Onk++04,Lau++07}.

It is important to notice that the formulae \mgc, \mgb, \mga, \msh,
and \msa\ have been calibrated on the isotropic spherical virial
coefficient (f=3/4 in the notation of Netzer, 1990), \mm\, and \mk\,
adopt f=1, and \mt, \mw, \mv, \mvb, and \mn\ are based on the
recalibration of the virial coefficient given by
\citet{Onk++04}\footnote{It is also generally assumed that
FWHM=2$\sigma_{\rm line}$. This is consistent on average with our
result for the Balmer lines, albeit with large scatter, but not for
the \ion{Mg}{2} line.}. Therefore, we expect the first set of \mbh\
estimators to give lower values than \mt\, by $\log 1.8 = 0.255$ dex,
and \mm\, and \mk\, to give lower values by $\log (1.8 \times 3/4) =
0.130$ dex.  In the case of \msh\, and \msa, an isotropic spherical
virial coefficient was used. However, their equations were based on a
different size-luminosity relation, hence making these formulae
approximately equivalent to having the same virial coefficient as \mt,
\mw, and \mn\ \citep{Sal++07}.  As noted by \cite{Tre++07}, \mw, \mn,
\msh, and \msa\ agree on average to within a few per cent of \mt.  The
more discrepant black hole masses are those with different virial coefficients,
which can differ on average by as much as
$0.38\pm0.05$ dex (i.e. more than a factor of two). Even renormalizing
these formulae to the same virial coefficient would still leave
discrepancies of order 0.1 dex: 
after renormalization, \mv\, and \mgb\ 
are approximately 0.1 dex larger than \mt\ while \mgc, \mm, and \mk\ are
still 0.1 dex smaller than \mt. This is approximately twice the
expected error on the mean given the size of the sample.

These results show that {\it systematic} errors as large as
$0.38\pm0.05$ dex can be introduced when comparing \mbh\, estimates
based on different diagnostics or when comparing AGN \mbh\, estimates
to local samples with direct \mbh\ measurements from stellar or
gaseous kinematics.

We conclude by discussing the absolute calibration of our fiducial
mass estimator. A study by \citet{Col++06} finds that for the local
sample with reverberation based \mbh, the widths of the Balmer lines
measured on mean spectra are systematically larger than those measured
on the r.m.s. spectra for variable objects, suggesting that a smaller
virial coefficient than that advocated by \citet{Onk++04} and used for
our fiducial mass estimator, should be adopted when measuring widths
from mean spectra. This could possibly indicate that our fiducial
\mbh\, are overestimated by approximately 0.15 dex. To investigate
this effect, in Section~\ref{sec:res} we apply our calibrated recipes
-- based on our fiducial estimator -- to the local sample with
reverberation based \mbh, using published line widths and fluxes
measured from single epoch and mean spectra.  As discussed in the next
section, we find an excellent agreement with the mass inferred from
reverberation mapping based on line widths from the r.m.s. spectra,
indicating that no such correction is necessary. Data for a larger
number of objects with reverberation-based masses are needed to
determine the zero point of the virial scalings more accurately, as
discussed in the next Section.

\begin{figure*}
\epsscale{1.0}
\plotone{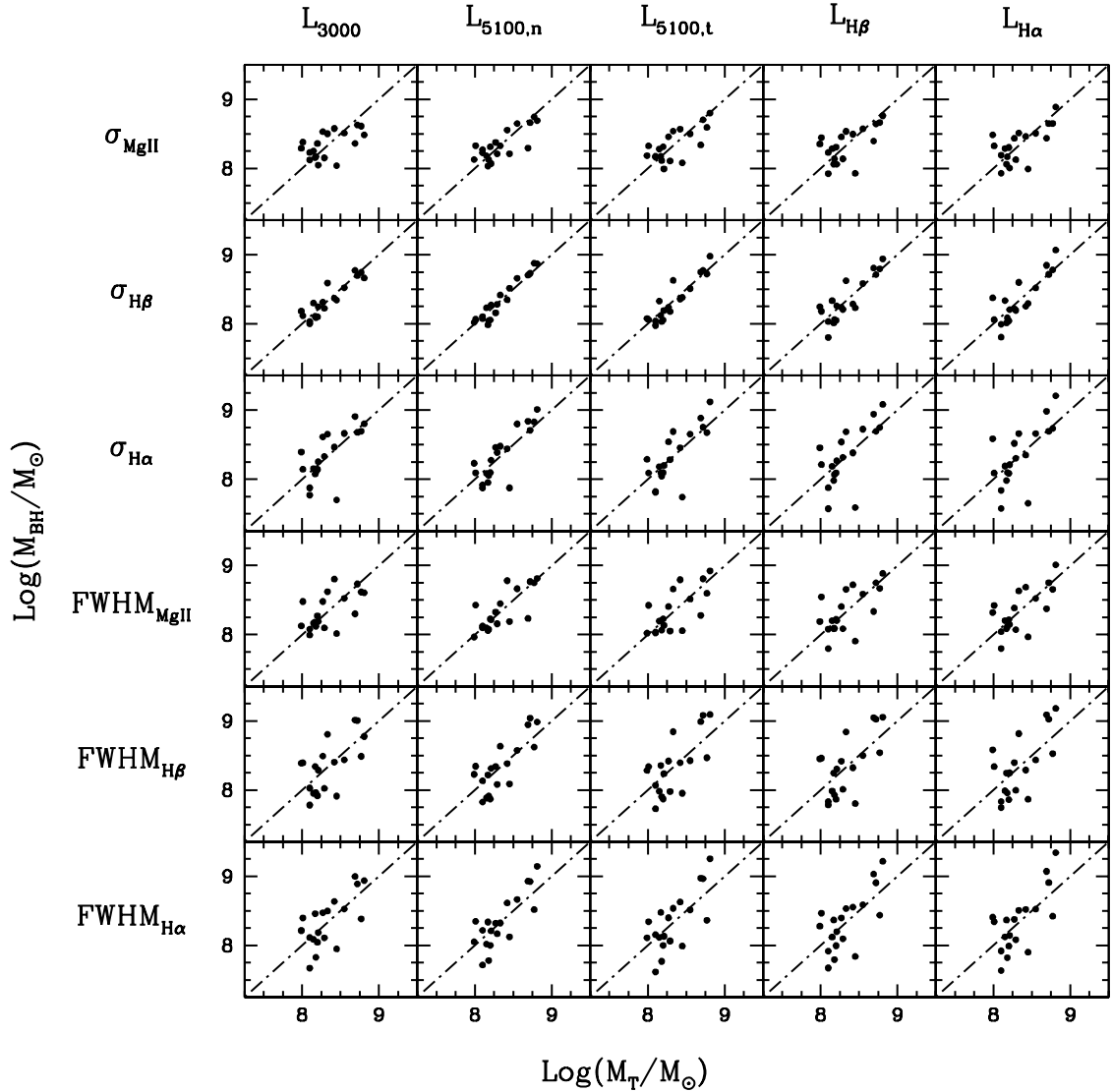} 
\caption{Comparison of \mbh\ estimates according to the new
cross-calibrated formulae as discussed in Section~\ref{sec:res}. A
relation of the form $\log$ \mbh $=\alpha + 2 \log {\rm v_{1000}} +
\beta \log {\rm L}$ is assumed, where ${\rm v_{1000}}$ is a velocity
estimator in units of 1000 \kms, and L is a luminosity estimator in
units of 10$^{44}$ erg s$^{-1}$ or 10$^{42}$ erg s$^{-1}$,
respectively for continuum or line luminosity. The slope $\beta$ is fixed at
0.47 for $L_{3000}$, 0.518 for $L_{5100,n}$, 0.69 for $L_{5100,t}$,
0.56 for $L_{\rm H\beta}$, and 0.55 for L$_{\rm H\alpha}$ as taken
from the literature. The best fit coefficients $\alpha$ are given in
Table~\ref{tab:alphas} together with the r.m.s. scatter of the
comparison. The typical measurement error bar is 0.05 dex.}
\label{fig:bhm}
\end{figure*}

\section{A set of self-consistent recipes}
\label{sec:res}

In this Section we examine all possible combinations of velocity and
flux estimators to produce a set of cross calibrated recipes. By
comparing with our fiducial mass estimator \mt, we compute the r.m.s.
scatter of the residuals to infer a lower limit on the intrinsic
uncertainty of each recipe.

In practice, we adopt the following relation:

\begin{equation}
\log M_{\rm BH} =\alpha + 2  \log {\rm v_{1000}} + \beta \log {\rm L},
\end{equation}

\noindent
where ${\rm v_{1000}}$ is a velocity estimator in units of 1000 \kms,
L is a luminosity estimator in units of 10$^{44}$ erg s$^{-1}$ or
10$^{42}$ erg s$^{-1}$, respectively for continuum or line luminosity,
and we find the $\alpha$ that best matches the \mt\, fiducial
estimates.  Our range in luminosities is too small to fit for $\beta$
as well, and therefore, we assert the following fiducial values:
$\beta=0.47$ for L$_{3000}$, 0.518 for L$_{5100,n}$, 0.67 for
L$_{5100,t}$, 0.56 for H$\beta$, and 0.55 for H$\alpha$. These choices
are based on the most current calibration of the size-luminosity
relation for each wavelength/line. In particular, following the
results of the study by \cite{Ben++06a} we adopt 0.518 as the slope of
the broad-line region size vs {\it nuclear} luminosity relation, while
for the size vs {\it total} luminosity (i.e. including host galaxy
contamination within the spectroscopic aperture) relation we adopt the
slope given by \citet{Kas++05}. Although the former slope is to be
preferred when the nuclear luminosity is available, we also provide
results for the second slope, which appears to be the best estimate
whenever the light from the nucleus and from the host galaxy cannot be
disentangled. In general, extrapolations well outside the range
considered here are to be done with caution, since most of the local
calibrators for the size-luminosity relation are Seyferts and PG
quasars in $10^{42} < L_{5100}({\rm erg s}^{-1}) < 10^{46}$.

We emphasize that this procedure produces self consistent mass
estimates, but these all share a common uncertainty in the zero
point. In practice, all $\alpha$ can be shifted by a constant if it
turns out that a different value of the virial coefficient for the
local sample of reverberation mapped AGNs is to be preferred. As a
sanity check, we applied our calibrated recipes to the local sample
with reverberation \mbh\, (\citet{Pet++04} and
recent updates by \citet{Ben++06b}, \citet{Den++06}, and
\citet{Ben++07}).  Based on the single epoch measurements given by
\citet{V+P06} we compared our two estimators based on the FWHM of \hb,
and on L$_{5100,t}$ and \hb\, line flux, with the reverberation
masses. The agreement is excellent, with average $\Delta \log {\rm
M}_{\rm BH,rev} - \Delta \log {\rm M}_{\rm BH}$ equal to 0.009 dex and
0.023, respectively, with r.m.s. scatter of 0.46 dex and 0.49
dex. Based on the measurements from mean spectra given by
\citet{Col++06} we compared our two estimators based on the FWHM and
line dispersion of \hb, and on L$_{5100,t}$, with the reverberation
masses. The agreement is again excellent, with average $\Delta \log
{\rm M}_{\rm BH,rev} - \Delta \log {\rm M}_{\rm BH}$ equal to 0.004
dex and 0.018, respectively, with r.m.s. scatter of 0.35 and 0.29 dex.
The scatter of the difference is combination of effects from uncertainties in
time-lag, line width, and luminosity measurements, and scatter in 
the size-luminosity relation.

The best fit values of $\alpha$ together with the r.m.s. scatter are
listed in Table~\ref{tab:alphas}. The best fit relations are shown in
Figure~\ref{fig:bhm}. We note that we included in the recipes the
combination of H$\beta$ \sline\, and L$_{5100,n}$ that was used to
compute \mt\, by \cite{Tre++07}. The goal of this exercise is to
estimate the measurement errors associated with the relation, given
that the input parameters were measured independently for this paper,
based on the Gauss-Hermite polynomial expansion fit. Since measurement
errors such as narrow line and continuum subtraction dominate over
pixel noise, given the high S/N of the Keck data, the r.m.s. of these
residuals divided by $\sqrt2$ is effectively the total measurement
error, i.e. $0.03$ dex. Thus we can conclude that, for all practical
purposes, the r.m.s. scatter that we observe for the other recipes is
intrinsic scatter in the relation and not measurement error. This is
also the rationale for not showing measurement error bars in the
plots.

Looking at Table~\ref{tab:alphas}, it appears that the smallest
relative scatter is obtained when comparing \mt\, to black hole masses
based on the same velocity scale $\sigma_{\rm H\beta}$. This is not
surprising, as the velocity scale enters with the square in the \mbh\
estimate and we have seen that velocity scales have typical relative
scatter of 0.1 dex. However, the scatter means that adopting
optical/UV continuum luminosity as a proxy of the size of the
broad-line region introduces an uncertainty of $\sim$ 0.10-0.15 dex in
the \mbh\ estimate, with the larger r.m.s. for the line luminosities.
The velocity scale that best matches the \sline\, of \hb\, is the line
width of MgII, which gives an r.m.s. scatter of 0.17 dex. This is
better than the \sline\, of \ha\, (0.20 dex) and the FWHM of \hb\,
(0.23 dex) -- as expected from the large distribution of FWHM/\sline\,
ratios -- which are in turn slightly better than the other
indicators. The worst match is obtained for line luminosities with
FWHM, with an r.m.s. scatter of $\sim0.24-0.33$ dex.  From this study,
we infer a lower limit to the relative accuracy of the various
indicators of order 0.1-0.2 dex, depending on the choice of
estimators. If the relationships presented here were to be extended
beyond the range of \mbh\ considered, it is likely that the scatter
will increase, as suggested by the slope of the points in the
corresponding panels.  Otherwise, the slope $\beta$ of the size
luminosity relation will have to be fitted independently.

\section{Summary}
\label{sec:sum}

In this paper we have used Keck and SDSS spectra of nineteen Seyferts
at $z=0.36$, to perform a comprehensive study of ``virial''
black hole mass estimators for broad line AGNs.  The main results can be
summarized as follows:

\begin{enumerate}

\item We have fit Gauss-Hermite series to the data in order to
measure the FWHM and \sline\, of MgII, \hb\,, and \ha\,, as well as
\ha\, and \hb\, luminosities and continuum luminosities at 3000\AA\
and 5100\AA.  Measurement errors are approximately 0.02 dex on the
MgII and \hb\ line widths, 0.04-0.05 on the \ha\ line widths, 0.01 dex
on the continuum luminosity, 0.01 dex on the \hb\ luminosity, and 0.06
dex on the \ha\ luminosity.

\item We have compared twelve formulae taken from the literature,
showing that \mbh\ estimates can differ systematically by as much as
$0.38\pm0.05$ dex (or $0.13\pm0.05$ dex, if the same virial
coefficient is adopted).  Such differences should be taken into
account when comparing data obtained with different methods.

\item We have cross-calibrated a set of 30 empirical recipes based on
all combinations of the velocity and luminosity indicators
corresponding to the \ion{Mg}{2}, \hb, and \ha\ broad lines.  Taking
the masses measured by Treu et al. (2007) as our fiducial black hole
masses, we find that: the absolute scale of the different indicators
is calibrated to within $\sim$0.05 dex; the best agreement is found
when using the line dispersion of \hb\, as a velocity estimator, with the
residual 0.1 dex r.m.s. scatter resulting from the various continuum luminosity
estimators; adopting the line dispersion of \ion{Mg}{2} raises the scatter
to 0.2 dex; for the other estimators the intrinsic scatter is in the
range 0.2-0.38 dex. This implies a lower limit of 0.1-0.2 dex on the
validity of each estimator for each individual case.

\end{enumerate}

The newly calibrated recipes should be useful to reduce the
sources of systematic uncertainties when comparing different studies.

\acknowledgments

This work is based on data obtained with the Hubble Space Telescope --
obtained at the Space Telescope Science Institute, which is operated
by the Association of Universities for Research in Astronomy, Inc.,
under NASA contract NAS 5-26555. These observations are associated
with program 10216 --, and with the 10m W.M. Keck Telescope, which is
operated as a scientific partnership among the California Institute of
Technology, the University of California and the National Aeronautics
and Space Administration. The Observatory was made possible by the
generous financial support of the W.M. Keck Foundation. The authors
wish to recognize and acknowledge the very significant cultural role
and reverence that the summit of Mauna Kea has always had within the
indigenous Hawaiian community.  We are most fortunate to have the
opportunity to conduct observations from this mountain. We acknowledge
financial support from NASA through HST grant GO-10216 and AR-10986.
TT acknowledges support from the NSF through CAREER award NSF-0642621,
and from the Sloan Foundation through a Sloan Research Fellowship.  
We thank Todd Boroson, Ross McLure, and Marianne Vestergaard for providing nuclear Fe
emission templates, and Brad Peterson for providing Table 1 in the paper by Collin et al. (2006) in electronic format.
We thank the referee for a careful report which improved the manuscript.


\begin{thebibliography}{38}
\expandafter\ifx\csname natexlab\endcsname\relax\def\natexlab#1{#1}\fi

\bibitem[{{Bentz} {et~al.}(2006){Bentz}, {Peterson}, {Pogge}, {Vestergaard}, \&
  {Onken}}]{Ben++06a}
{Bentz}, M.~C., {Peterson}, B.~M., {Pogge}, R.~W., {Vestergaard}, M., \&
  {Onken}, C.~A. 2006, \apj, 644, 133

\bibitem[{{Bentz} {et~al.}(2006)}]{Ben++06b} {Bentz}, M.~C.,
et~al. 2006, \apj, 651, 775

\bibitem[{{Bentz} {et~al.}(2007)}]{Ben++07} {Bentz}, M.~C.,
et~al. 2007, \apj, 662, 205

\bibitem[{{Cardelli} {et~al.}(1989){Cardelli}, {Clayton}, \& {Mathis}}]{CCM89}
{Cardelli}, J.~A., {Clayton}, G.~C., \& {Mathis}, J.~S. 1989, \apj, 345, 245

\bibitem[{{Collin} {et~al.}(2006){Collin}, {Kawaguchi}, {Peterson}, \&
  {Vestergaard}}]{Col++06}
{Collin}, S., {Kawaguchi}, T., {Peterson}, B.~M., \& {Vestergaard}, M. 2006,
  \aap, 456, 75

\bibitem[{{Croton} {et~al.}(2006){Croton}, {Springel}, {White}, {De Lucia},
  {Frenk}, {Gao}, {Jenkins}, {Kauffmann}, {Navarro}, \& {Yoshida}}]{Cro++06b}
{Croton}, D.~J., {Springel}, V., {White}, S.~D.~M., {De Lucia}, G., {Frenk},
  C.~S., {Gao}, L., {Jenkins}, A., {Kauffmann}, G., {Navarro}, J.~F., \&
  {Yoshida}, N. 2006, \mnras, 365, 11

\bibitem[{{Denney} {et~al.}(2006)}]{Den++06} {Denney}, K.~D.,
et~al. 2006, \apj, 653, 152

\bibitem[{{Di Matteo} {et~al.}(2005){Di Matteo}, {Springel}, \&
  {Hernquist}}]{DMS05}
{Di Matteo}, T., {Springel}, V., \& {Hernquist}, L. 2005, \nat, 433, 604

\bibitem[{{Ferrarese} \& {Ford}(2005)}]{F+F05}
{Ferrarese}, L. \& {Ford}, H. 2005, Space Science Reviews, 116, 523

\bibitem[{{Gerhard}(1993)}]{Ger93}
{Gerhard}, O.~E. 1993, \mnras, 265, 213

\bibitem[{{Greene} \& {Ho}(2005)}]{G+H05}
{Greene}, J.~E. \& {Ho}, L.~C. 2005, \apj, 630, 122

\bibitem[{{Kaspi} {et~al.}(2005){Kaspi}, {Maoz}, {Netzer}, {Peterson},
  {Vestergaard}, \& {Jannuzi}}]{Kas++05}
{Kaspi}, S., {Maoz}, D., {Netzer}, H., {Peterson}, B.~M., {Vestergaard}, M., \&
  {Jannuzi}, B.~T. 2005, \apj, 629, 61

\bibitem[{{Kaspi} {et~al.}(2000){Kaspi}, {Smith}, {Netzer}, {Maoz}, {Jannuzi},
  \& {Giveon}}]{Kas++00}
{Kaspi}, S., {Smith}, P.~S., {Netzer}, H., {Maoz}, D., {Jannuzi}, B.~T., \&
  {Giveon}, U. 2000, \apj, 533, 631

\bibitem[{{Kollmeier} {et~al.}(2006){Kollmeier}, {Onken}, {Kochanek}, {Gould},
  {Weinberg}, {Dietrich}, {Cool}, {Dey}, {Eisenstein}, {Jannuzi}, {Le Floc'h},
  \& {Stern}}]{Kol++06}
{Kollmeier}, J.~A., {Onken}, C.~A., {Kochanek}, C.~S., {Gould}, A., {Weinberg},
  D.~H., {Dietrich}, M., {Cool}, R., {Dey}, A., {Eisenstein}, D.~J., {Jannuzi},
  B.~T., {Le Floc'h}, E., \& {Stern}, D. 2006, \apj, 648, 128

\bibitem[{{Kormendy} \& {Gebhardt}(2001)}]{K+G01}
{Kormendy}, J. \& {Gebhardt}, K. 2001, 586, 363

\bibitem[{{Lacy} {et~al.}(1982){Lacy}, {Malkan}, {Becklin}, {Soifer},
  {Neugebauer}, {Matthews}, {Wu}, {Boggess}, \& {Gull}}]{Lac++82}
{Lacy}, J.~H., {Malkan}, M., {Becklin}, E.~E., {Soifer}, B.~T., {Neugebauer},
  G., {Matthews}, K., {Wu}, C.-C., {Boggess}, A., \& {Gull}, T.~R. 1982, \apj,
  256, 75

\bibitem[{{Lauer} {et~al.}(2007){Lauer}, {Tremaine}, {Richstone}, \&
  {Faber}}]{Lau++07}
{Lauer}, T.~R., {Tremaine}, S., {Richstone}, D., \& {Faber}, S.~M. 2007, ArXiv
  e-prints, 0705.4103

\bibitem[{{Malkan}(1983)}]{Mal83}
{Malkan}, M.~A. 1983, \apjl, 264, L1

\bibitem[{{McLure} \& {Jarvis}(2002)}]{M+J02}
{McLure}, R.~J. \& {Jarvis}, M.~J. 2002, \mnras, 337, 109

\bibitem[{{Netzer} \& {Trakhtenbrot}(2007)}]{N+T07}
{Netzer}, H. \& {Trakhtenbrot}, B. 2007, \apj, 654, 754

\bibitem[{{Oke} {et~al.}(1995){Oke}, {Cohen}, {Carr}, {Cromer}, {Dingizian},
  {Harris}, {Labrecque}, {Lucinio}, {Schaal}, {Epps}, \& {Miller}}]{Oke++95}
{Oke}, J.~B., {Cohen}, J.~G., {Carr}, M., {Cromer}, J., {Dingizian}, A.,
  {Harris}, F.~H., {Labrecque}, S., {Lucinio}, R., {Schaal}, W., {Epps}, H., \&
  {Miller}, J. 1995, \pasp, 107, 375

\bibitem[{{Onken} {et~al.}(2004){Onken}, {Ferrarese}, {Merritt}, {Peterson},
  {Pogge}, {Vestergaard}, \& {Wandel}}]{Onk++04}
{Onken}, C.~A., {Ferrarese}, L., {Merritt}, D., {Peterson}, B.~M., {Pogge},
  R.~W., {Vestergaard}, M., \& {Wandel}, A. 2004, \apj, 615, 645

\bibitem[{{Osterbrock}(1989)}]{Ost89}
{Osterbrock}, D.~E. 1989, {Astrophysics of gaseous nebulae and active galactic
  nuclei} University Science Books, 1989, 422 p.

\bibitem[{{Peterson} {et~al.}(2004){Peterson}, {Ferrarese}, {Gilbert}, {Kaspi},
  {Malkan}, {Maoz}, {Merritt}, {Netzer}, {Onken}, {Pogge}, {Vestergaard}, \&
  {Wandel}}]{Pet++04}
{Peterson}, B.~M., {Ferrarese}, L., {Gilbert}, K.~M., {Kaspi}, S., {Malkan},
  M.~A., {Maoz}, D., {Merritt}, D., {Netzer}, H., {Onken}, C.~A., {Pogge},
  R.~W., {Vestergaard}, M., \& {Wandel}, A. 2004, \apj, 613, 682

\bibitem[{{Salviander} {et~al.}(2007){Salviander}, {Shields}, {Gebhardt}, \&
  {Bonning}}]{Sal++07}
{Salviander}, S., {Shields}, G.~A., {Gebhardt}, K., \& {Bonning}, E.~W. 2007,
  \apj, 662, 131

\bibitem[{{Shields} {et~al.}(2003){Shields}, {Gebhardt}, {Salviander}, {Wills},
  {Xie}, {Brotherton}, {Yuan}, \& {Dietrich}}]{Shi++03}
{Shields}, G.~A., {Gebhardt}, K., {Salviander}, S., {Wills}, B.~J., {Xie}, B.,
  {Brotherton}, M.~S., {Yuan}, J., \& {Dietrich}, M. 2003, \apj, 583, 124

\bibitem[{{Shields} {et~al.}(2006){Shields}, {Menezes}, {Massart}, \& {Vanden
  Bout}}]{Shi++06}
{Shields}, G.~A., {Menezes}, K.~L., {Massart}, C.~A., \& {Vanden Bout}, P.
  2006, \apj, 641, 683

\bibitem[{{Shuder}(1984)}]{Shu84}
{Shuder}, J.~M. 1984, \apj, 280, 491

\bibitem[{{Treu} {et~al.}(2004){Treu}, {Malkan}, \& {Blandford}}]{TMB04}
{Treu}, T., {Malkan}, M.~A., \& {Blandford}, R.~D. 2004, \apjl, 615, L97

\bibitem[{{Treu} {et~al.}(2007){Treu}, {Woo}, {Malkan}, \&
  {Blandford}}]{Tre++07}
{Treu}, T., {Woo}, J.-H., {Malkan}, M.~A., \& {Blandford}, R.~D. 2007, ApJ, 667, 117

\bibitem[{{van der Marel} \& {Franx}(1993)}]{v+F93}
{van der Marel}, R.~P. \& {Franx}, M. 1993, \apj, 407, 525


\bibitem[{{Vestergaard} \& {Peterson}(2006)}]{V+P06}
{Vestergaard}, M. \& {Peterson}, B.~M. 2006, \apj, 641, 689

\bibitem[{{Wandel} {et~al.}(1999){Wandel}, {Peterson}, \& {Malkan}}]{WPM99}
{Wandel}, A., {Peterson}, B.~M., \& {Malkan}, M.~A. 1999, \apj, 526, 579

\bibitem[{{Webb} \& {Malkan}(2000)}]{W+M00}
{Webb}, W. \& {Malkan}, M. 2000, \apj, 540, 652

\bibitem[{{Woo} {et~al.}(2006){Woo}, {Treu}, {Malkan}, \&
  {Blandford}}]{Woo++06}
{Woo}, J.-H., {Treu}, T., {Malkan}, M.~A., \& {Blandford}, R.~D. 2006, \apj,
  645, 900

\bibitem[{{Woo} {et~al.}(2007){Woo}, {Treu}, {Malkan}, {Ferry}, \&
  {Misch}}]{Woo++07}
{Woo}, J.-H., {Treu}, T., {Malkan}, M.~A., {Ferry}, M.~A., \& {Misch}, T. 2007,
  \apj, 661, 60

\bibitem[{{Woo} \& {Urry}(2002{\natexlab{a}})}]{W+U02b}
{Woo}, J.-H. \& {Urry}, C.~M. 2002{\natexlab{a}}, \apj, 579, 530

\bibitem[{{Woo} \& {Urry}(2002{\natexlab{b}})}]{W+U02a}
---. 2002{\natexlab{b}}, \apjl, 581, L5

\end{thebibliography}

\begin{deluxetable}{lrrcrrr}
\tablewidth{0pt}
\tablecaption{Sample properties}
\tablehead{
\colhead{Name}    &
\colhead{RA (J2000)}     &
\colhead{DEC (J2000)}    &
\colhead{z}    &
\colhead{i'}
\\
\colhead{(1)} &
\colhead{(2)} &
\colhead{(3)} &
\colhead{(4)} &
\colhead{(5)} &}
\tablecolumns{5}
\startdata
S01 & 15 39 16.23 & +03 23 22.06 & 0.3592 & 18.74 \\
S02 & 16 11 11.67 & +51 31 31.12 & 0.3544 & 18.94 \\
S03 & 17 32 03.11 & +61 17 51.96 & 0.3588 & 18.20 \\
S04 & 21 02 11.51 & -06 46 45.03 & 0.3578 & 18.41 \\
S05 & 21 04 51.85 & -07 12 09.45 & 0.3530 & 18.35 \\
S06 & 21 20 34.19 & -06 41 22.24 & 0.3684 & 18.41 \\
S07 & 23 09 46.14 & +00 00 48.91 & 0.3518 & 18.11 \\
S08 & 23 59 53.44 & -09 36 55.53 & 0.3585 & 18.43 \\
S09 & 00 59 16.11 & +15 38 16.08 & 0.3542 & 18.16 \\
S10 & 01 01 12.07 & -09 45 00.76 & 0.3506 & 17.92 \\
S11 & 01 07 15.97 & -08 34 29.40 & 0.3557 & 18.34 \\
S12 & 02 13 40.60 & +13 47 56.06 & 0.3575 & 18.12 \\
S21 & 11 05 56.18 & +03 12 43.26 & 0.3534 & 17.21 \\
S23 & 14 00 16.66 & -01 08 22.19 & 0.3510 &  18.08 \\
S24 & 14 00 34.71 & +00 47 33.48 & 0.3615 & 18.21 \\
S26 & 15 29 22.26 & +59 28 54.56 & 0.3691 & 18.88 \\
S27 & 15 36 51.28 & +54 14 42.71 & 0.3667 & 18.80 \\
S28 & 16 11 56.30 & +45 16 11.04 & 0.3660 & 18.59 \\
S29 & 21 58 41.93 & -01 15 00.33 & 0.3576 & 18.77 \\
\enddata
\label{tab:sample}
\tablecomments{
Col. (1): Target ID.  
Col. (2): RA.
Col. (3): DEC.
Col. (4): Redshift from SDSS DR4.       
Col. (5): Extinction corrected $i'$ AB magnitude from SDSS photometry.}
\end{deluxetable}

\begin{deluxetable}{lccccccccccccc}
\tablewidth{0pt}
\tablecaption{Measured Properties}
\tablehead{
\colhead{Name}    &
\colhead{$\sigma_{\rm MgII}$}    &
\colhead{$\sigma_{\rm H\beta}$}    &
\colhead{$\sigma_{\rm H\alpha}$}    &
\colhead{$\rm FWHM_{\rm MgII}$}    &
\colhead{$\rm FWHM_{\rm H\beta}$}    &
\colhead{$\rm FWHM_{\rm H\alpha}$}    &  
\colhead{$L_{3000}$}    &
\colhead{$L_{5100,n}$}    &
\colhead{$L_{5100,t}$}    &
\colhead{$L_{{\rm H}\beta}$}    &
\colhead{$L_{{\rm H}\alpha}$}  &
\colhead{$f_{{\rm H}\beta,[OIII]}$} &
\colhead{$f_{{\rm H}\beta,nt}$} \\
\colhead{(1)} &
\colhead{(2)} &
\colhead{(3)} &
\colhead{(4)} &
\colhead{(5)} &
\colhead{(6)} &
\colhead{(7)} &
\colhead{(8)} &
\colhead{(9)} &
\colhead{(10)} &
\colhead{(11)} &
\colhead{(12)} &
\colhead{(13)} &
\colhead{(14)} }
\tablecolumns{14}
\startdata
S01 & 2260. & 2133. & 1847. & 4418. & 4755. & 3420. &  1.91 &  0.74 &  1.66 &  2.29 &  7.24 & 0.1  & 0.05 \\
S02 & 2914. & 1928. & 2113. & 4000. & 5188. & 3442. &  2.14 &  0.36 &  1.51 &  3.09 & 21.38 & 0.17 & 0.10 \\
S03 & 2147. & 1745. & 1698. & 3345. & 2945. & 2634. &  3.98 &  1.69 &  2.82 &  3.80 & 14.45 & 0.13 & 0.04  \\
S04 & 2253. & 2392. &  983. & 3636. & 3100. & 2617. &  1.86 &  1.42 &  2.24 &  1.35 &  6.92 & 0.08 & 0.06  \\
S05 & 3784. & 3297. & 2686. & 6315. & 5220. & 3751. &  3.39 &  2.04 &  2.75 &  4.37 & 16.60 & 0.11 & 0.05  \\
S06 & 2353. & 1664. & 1382. & 4023. & 4625. & 4306. &  2.75 &  0.54 &  2.69 &  2.00 &  7.94 & 0.1 &  0.13 \\
S07 & 3297. & 2500. & 2531. & 5561. & 4815. & 4326. &  3.72 &  2.26 &  3.02 &  4.90 & 15.14 & 0.1 &  0.06 \\
S08 & 2365. & 1538. & 1015. & 3380. & 3372. & 3017. &  2.29 &  1.25 &  2.57 &  1.12 &  4.47 & 0.1 &  0.08 \\
S09 & 2542. & 2013. & 1462. & 3824. & 2865. & 2710. &  3.16 &  0.78 &  3.09 &  3.80 & 15.49 & 0.11 & 0.04 \\
S10 & 2897. & 1690. & 2056. & 4532. & 4410. & 3498. &  7.08 &  1.11 &  3.80 &  4.79 & 17.78 & 0.11 & 0.02 \\
S11 & 2858. & 1590. & 1435. & 4291. & 2733. & 2569. &  3.24 &  0.88 &  2.45 &  2.75 & 10.72 & 0.08 & 0.01 \\
S12 & 2672. & 3213. & 3232. & 4132. & 9005. & 7163. &  4.37 &  1.05 &  3.24 &  5.01 & 23.99 & 0.06 & 0.02 \\
S21 & 3450. & 3172. & 3208. & 6582. & 7681. & 7493. &  2.69 &  2.30 &  7.24 &  9.12 & 63.10 & 0.07 & 0.03  \\
S23 & 3941. & 3196. & 2688. & 7378. & 9700. & 6870. &  3.09 &  1.20 &  3.55 &  3.47 & 14.13 & 0.1  & 0.05 \\
S24 & 3480. & 2886. & 2667. & 6628. & 7864. & 4468. &  2.82 &  0.44 &  2.95 &  3.47 & 12.59 & 0.1  & 0.04  \\
S26 & 3370. & 1862. & 1657. & 6265. & 5451. & 4440. &  1.78 &  0.50 &  1.58 &  2.69 &  6.46 & 0.07 & 0.12  \\
S27 & 2699. & 1609. & 1157. & 3766. & 2567. & 1832. &  2.19 &  0.92 &  1.82 &  2.45 &  8.32 & 0.17 & 0.05  \\
S28 & 3926. & 2313. & 2221. & 8436. & 5116. & 5412. &  2.45 &  0.76 &  2.29 &  1.91 &  6.61 & 0.09 & 0.03  \\
S29 & 2556. & 1744. & 1520. & 4003. & 3190. & 2216. &  2.09 &  0.59 &  1.74 &  2.04 &  9.12 & 0.1  & 0.05  \\
\enddata
\label{tab:derive}
\tablecomments{
Col. (1): Target ID.  
Col. (2): \sline\, of MgII (\kms) measured on the line model fit to Keck data. The average error is 0.017 dex.
Col. (3): \sline\, of \hb\ (\kms) measured on the line model fit to Keck data. The average error is 0.017 dex. 
Col. (4): \sline\, of \ha\ (\kms) measured on the line model fit to SDSS data. The average error is  0.051 dex.
Col. (5): FWHM of MgII (\kms) measured on the line model fit to Keck data. The average error is 0.017 dex.
Col. (6): FWHM of \hb\ (\kms) measured on the line model fit to Keck data. The average error is 0.017 dex.     
Col. (7): FWHM of \ha\ (\kms) measured on the line model fit to SDSS data. The average error is 0.040 dex. 
Col. (8): Rest frame luminosity at 3000 $\rm \AA$ in 10$^{44}$ erg s$^{-1}$. The average error is 0.014 dex. The actual error 
will be dominated by variability of order 10\%, as for all luminosities listed in this Table.
Col. (9): Rest frame nuclear luminosity at 5100 $\rm \AA$ in 10$^{44}$ erg s$^{-1}$, from Treu et al.\ (2007).  The average error is 0.08 dex.
Col. (10): Rest frame total luminosity at 5100 $\rm \AA$ in 10$^{44}$ erg s$^{-1}$. The average error is 0.014 dex. 
Col. (11): Rest frame \hb\ line luminosity in 10$^{42}$ erg s$^{-1}$. The average error is 0.011 dex.
Col. (12): Rest frame \ha\ line luminosity in 10$^{42}$ erg s$^{-1}$. The average error is 0.062 dex.
Col. (13): Narrow component of \hb\ to [OIII]$\lambda$5007 flux ratio.
Col. (14): Fraction of flux of the \hb\ line in the narrow component.
}  
\end{deluxetable}

\begin{deluxetable}{lrcccc}
\tablewidth{0pt}
\tablecaption{M$_{\rm BH}$ Estimator Factors}
\tablehead{
\colhead{         }    &
\colhead{$L_{3000}$}    &
\colhead{$L_{5100,n}$}    &
\colhead{$L_{5100,t}$}    &
\colhead{$L_{\rm H\beta}$}    &
\colhead{$L_{\rm H\alpha}$}    
\\
\colhead{        } &
\colhead{(0.47)} &
\colhead{(0.518)} &
\colhead{(0.69)} &
\colhead{(0.56)} &
\colhead{(0.55)} } 
\tablecolumns{6}
\startdata
$\sigma_{\rm MgII}$      & 7.207$\pm$   0.052, 0.219 &  7.429$\pm$   0.039, 0.166 & 7.133$\pm$   0.045, 0.191 & 7.150$\pm$   0.053, 0.226 & 6.824$\pm$   0.051, 0.218 \\
$\sigma_{\rm H\beta}$    & 7.458$\pm$   0.027, 0.112 &  7.680$\pm$   0.021, 0.090 & 7.383$\pm$   0.028, 0.119 & 7.401$\pm$   0.038, 0.163 & 7.074$\pm$   0.042, 0.178 \\
$\sigma_{\rm H\alpha}$   & 7.588$\pm$   0.061, 0.260 &  7.810$\pm$   0.048, 0.205 & 7.514$\pm$   0.060, 0.253 & 7.532$\pm$   0.074, 0.313 & 7.205$\pm$   0.075, 0.319 \\
$\rm FWHM_{\rm MgII}$    & 6.767$\pm$   0.055, 0.233 &  6.990$\pm$   0.045, 0.191 & 6.693$\pm$   0.053, 0.224 & 6.711$\pm$   0.059, 0.251 & 6.384$\pm$   0.056, 0.236 \\
$\rm FWHM_{\rm H\beta}$  & 6.803$\pm$   0.069, 0.292 &  7.026$\pm$   0.056, 0.235 & 6.729$\pm$   0.071, 0.300 & 6.747$\pm$   0.077, 0.327 & 6.420$\pm$   0.079, 0.335 \\
$\rm FWHM_{\rm H\alpha}$ & 6.986$\pm$   0.064, 0.270 &  7.209$\pm$   0.055, 0.233 & 6.912$\pm$   0.069, 0.293 & 6.930$\pm$   0.071, 0.303 & 6.603$\pm$   0.073, 0.311 \\
\enddata
\label{tab:alphas}

\tablecomments{Normalization constants for $M_{\rm BH}$.  Entries are 
$\alpha \pm$ error, r.m.s. scatter of $\log$ M$_{\rm BH}$ vs. $\log$
M$_{\rm T}$ (Fig.~\ref{fig:bhm}).  The $\alpha$ coefficients are determined using
the general formula, $\log$ \mbh $=\alpha + 2\log {\rm v_{1000}} +
\beta \log {\rm L}$, where ${\rm v_{1000}}$ is the velocity estimator
in units of 1000~\kms, and L is the luminosity estimator, which is
divided by $10^{44} \mbox{ \rm erg s}^{-1}$ for continuum luminosity
measurements and by $10^{42} \mbox{
\rm erg s}^{-1}$ for line luminosity measurements.  The values used
for $\beta$ are listed below the luminosities.}
\end{deluxetable}

\end{document}